\documentclass[]{spie}

\title{\Large\bf Quantum Knots}

\author{Louis H. Kauffman\supit{a} and Samuel J. Lomonaco Jr.2\supit{b}
\skiplinehalf
\supit{a} Department of Mathematics, Statistics and Computer Science  
(m/c 249), 851 South Morgan Street, University of Illinois at Chicago,
Chicago, Illinois 60607-7045, USA \\
\supit{b} Department of Computer Science and Electrical Engineering, University of
Maryland Baltimore County, 1000 Hilltop Circle, Baltimore, MD 21250, USA}
 
\authorinfo{Further author information: L.H.K.  E-mail: kauffman@uic.edu,
S.J.L. Jr.: E-mail: lomonaco@umbc.edu}
 
\begin{document} 
  \maketitle 

\begin{abstract} 
This paper proposes the definition of a {\em quantum knot} as a linear superposition of classical knots in three dimensional space.
The definition is constructed and applications are discussed. Then the paper details extensions and also limitations of the Aravind Hypothesis
for comparing quantum measurement with classical topological measurement. We propose a separate, network model for quantum evolution and measurement, where the
background space is replaced by an evolving network. In this model there is an analog of the Aravind Hypothesis that promises to directly illuminate relationships
between physics, topology and quantum knots.
\end{abstract}


\section{INTRODUCTION}
We define a {\em quantum knot}  to be a linear superposition of classical knots in three dimensional space.
\bigbreak

The paper begins with a short discussion of the concept of knots and links. While there are many examples of this sort of topological pattern, we emphasize that 
it is indeed a mathematical pattern and has the possibility of matching models in many realms including the quantum realm. The second section of the paper 
gives the definition in more detail. The third section discusses a number of uses for the concept of a quantum knot.
In section 4 we discuss the Aravind Hypothesis for comparing quantum entanglement and topological entanglement, and compare it with a network model
for quantum measurement. The Aravind Hypothesis suggests modeling observation of a topological system (such as a link of curves in three dimensional space)
by an act of cutting and deleting one of the curves. Using this method one obtains analogies between topological observation and quantum measurement.
There are many discrepancies in this comparison. We detail our understanding of the situation in Section 5. In Section 6 we give a network model for quantum
measurement, and show that it involves the act of cutting an edge of the network, and inserting a density matrix at the edge. Thus there is an analogy between
this model for measurement and the Aravind Hypothesis. Sections 5 and 6 contain discussions of consequences of this point of view. The present paper uses
results of previous papers \cite{TEQE,Spie,BG} of the authors, but can be read independently of them.
\bigbreak

It has been our intent here to take a number of different points of view on the notion of a quantum knot as a superposition of topologies. One can look at this 
notion of quantum knot as possibly describing real systems such as vortices in superfluid helium, or small molecules tunneling among different bonded states.
One can also imagine notions of measurement for classical knots that might match the patterns of quantum states. And one can think of network models for 
worlds of interaction and try to see the relevance of topological structure to such models.
\bigbreak

\section{What is a Knot?}
A {\em knot} is an embedding of a circle into Euclidean $3$-space $R^{3}.$ A {\em} link is an embedding of a disjoint collection of circles into 
$R^{3}.$ Knots and links are studied up to the equivalence relation of {\em ambient isotopy}. Two embeddings are ambient isotopic if
one can be obtained from the other by a deformation through a family of embeddings. A knot is {\em knotted} if it is not equivalent to a flat circle in a plane.
A link is {\em linked} if it is not equivalent to a disjoint collection of flat circles.
\bigbreak

A {\em graphical mathematical model} allows one to represent knots and links by $4$-regular plane graphs with extra structure at the vertices.
These graphs with extra structure are called {\em knot and link diagrams}.  In the graphical model, there combinatorial moves that generate the analog of 
ambient isotopy (the Reidemeister moves as in Figure 1). A Theorem of Reidemeister assures us that two knots or links are ambient isotopic if and only if the
corresponding diagrams are equivalent by a finite sequence of these moves.
\bigbreak

{\tt    \setlength{\unitlength}{0.92pt}
\begin{picture}(368,280)
\thicklines   \put(234,59){Trefoil Diagram}
              \put(33,13){Reidemeister Moves}
              \put(303,121){\vector(0,-1){36}}
              \put(303,239){\vector(0,-1){105}}
              \put(259,240){\vector(1,0){43}}
              \put(258,205){\vector(0,1){36}}
              \put(309,199){\vector(1,0){47}}
              \put(302,85){\vector(-1,0){70}}
              \put(257,129){\vector(0,1){62}}
              \put(356,128){\vector(-1,0){99}}
              \put(356,198){\vector(0,-1){68}}
              \put(233,199){\vector(1,0){59}}
              \put(232,87){\vector(0,1){112}}
              \put(10,30){\framebox(191,240){}}
              \put(96,99){\vector(1,0){19}}
              \put(110,99){\vector(-1,0){17}}
              \put(73,173){\vector(1,0){19}}
              \put(87,173){\vector(-1,0){17}}
              \put(75,243){\vector(1,0){19}}
              \put(89,243){\vector(-1,0){17}}
              \put(193,80){\line(-1,-1){16}}
              \put(153,40){\line(1,1){16}}
              \put(153,40){\line(-1,1){15}}
              \put(113,80){\line(1,-1){18}}
              \put(115,118){\line(1,-1){79}}
              \put(115,39){\line(1,1){35}}
              \put(159,87){\line(1,1){34}}
              \put(96,82){\line(-1,1){15}}
              \put(55,122){\line(1,-1){17}}
              \put(40,107){\line(1,1){15}}
              \put(16,82){\line(1,1){16}}
              \put(61,88){\line(1,1){34}}
              \put(16,43){\line(1,1){35}}
              \put(16,122){\line(1,-1){79}}
              \put(133,174){\line(-1,-2){20}}
              \put(114,214){\line(1,-2){19}}
              \put(154,174){\line(1,-2){19}}
              \put(153,174){\line(1,2){20}}
              \put(45,153){\line(4,-3){24}}
              \put(20,175){\line(6,-5){16}}
              \put(36,187){\line(-4,-3){16}}
              \put(69,215){\line(-6,-5){24}}
              \put(56,173){\line(-1,-1){39}}
              \put(16,215){\line(1,-1){40}}
              \put(152,225){\line(-1,0){40}}
              \put(152,264){\line(0,-1){40}}
              \put(112,265){\line(1,0){40}}
              \put(29,243){\line(-1,-1){15}}
              \put(55,266){\line(-1,-1){16}}
              \put(55,226){\line(0,1){40}}
              \put(14,266){\line(1,-1){40}}
\end{picture}}

\begin{center}
{\bf Figure 1 - The Reidemeister Moves} \bigbreak
\end{center}

This completes a sketch of the mathematical concept of a knot and how it is represented in terms of both continuous and combinatorial structures.
Knotted phenomena (phenomena that can be modeled with this mathematical concept of a knot or link) occur in a wide variety of contexts: knotted rope,
woven clothing, the behaviour of telephone cords, the structure of the DNA molecule, the structure of long polymer chains, knotted trajectories in
dynamical systems, vortices in three dimensional fluids, small knotted molecules and even a species of eel that knots itself and slides the knot along its
body to clean itself (a biological use of self-reference).
\bigbreak

\section{What is a Quantum Knot?}
\begin{center}
{\tt    \setlength{\unitlength}{0.92pt}
\begin{picture}(176,264)
\thicklines   \put(127,188){\line(2,-1){15}}
              \put(127,188){\line(-3,-4){12}}
              \put(95,168){\vector(4,1){71}}
\thinlines    \put(30,220){\circle*{16}}
              \put(33,209){\circle*{16}}
              \put(39,228){\circle*{16}}
              \put(45,213){\circle*{16}}
              \put(37,214){\circle*{16}}
              \put(47,223){\circle*{16}}
\thicklines   \put(123,233){\circle*{10}}
              \put(134,234){\circle{40}}
              \put(115,230){\line(-4,-1){22}}
              \put(93,224){\line(1,0){26}}
              \put(133,215){\line(-3,-1){26}}
              \put(132,215){\line(5,-2){23}}
              \put(132,213){\line(-1,-5){5}}
              \put(66,216){$+$}
              \put(75,200){\vector(0,-1){64}}
              \put(135,116){\line(0,-1){92}}
              \put(32,108){\line(4,1){72}}
              \put(102,12){\line(-3,2){91}}
              \put(12,73){\line(3,1){84}}
              \put(111,107){\line(5,2){23}}
              \put(135,23){\line(-4,3){72}}
              \put(31,107){\line(6,-5){18}}
              \put(104,125){\line(0,-1){71}}
              \put(103,11){\line(0,1){28}}
              \put(123,219){\line(2,1){13}}
\end{picture}}

{\bf Figure 2 - Observing a Quantum Knot} \bigbreak
\end{center}

\noindent {\bf Definition.} A {\em quantum knot} is a linear superposition of classical knots.
\bigbreak

Figure 2 illustrates the notion that a quantum knot is an enigma of possible knots that resolves into particular topological structures when it is 
observed (measured).
\bigbreak

For example, we can let ${\bf K}$ stand for the collection of all knots, choosing one representative from each equivalence class.
This is a denumerable collection and we can form the formal infinite superposition of each of these knots with some appropriate
amplitude $\rho(K) e^{i\theta(K)}$ for each knot $K \in {\bf K},$ with $\rho(K)$ a non-negative real number.
$${\bf Q} = \Sigma_{K \in {\bf K}} \rho(K)e^{i\theta(K)} |K \rangle.$$
We assume that
$$\Sigma_{K \in {\bf K}} \rho(K)^{2} = 1.$$
${\bf Q}$ is the form of the most general quantum knot. Any particular quantum knot is obtained by specializing the associated 
amplitudes for the individual knots. A measurement of ${\bf Q}$ will yield the state $| K \rangle$ with probability $\rho(K)^2.$
\bigbreak

An example of a more restricted quantum knot can be obtained from a flat diagram such that there are two choices for over and under crossing
at each node of the diagram. Then we can make $2^N$ knot diagrams from the flat diagram and we can sum over representatives for the different
classes of knots that can be made from the given flat diagram. In this way, you can think of the flat diagram as representing a quantum knot
whose potential observed knots correspond to ways to resolve the crossings of the diagram. Or you could just superimpose a few random knots.
\bigbreak

\section{What are Some Possible Uses for Quantum Knots?}
\begin{enumerate}
\item The theory of vortices in supercooled Helium as proposed by Rasetti and Regge \cite{RR} uses the concept of quantum knot quite explicitly.
The vortex itself is a quantum phenomenon, and their theory uses a collection of observables that measure a planar curve (projection) of the knot, and then
other operators measure the over or under crossing structure of the nodes of this plane curve. It remains to be seen whether one can compute the multiplicity of 
possible knotted structures that are implicit in a given vortex.
\item In a knotted molecule there is some probability of {\em tunneling}, whose effect would be to change (from a given point of view) an under-crossing
to an over-crossing in the knotted structure. This is analogous to the way topology of large molecules such as DNA is changed by the presence of topological
enzymes that can cut a  bond, allow strand-passage and reseal the bond. But here we envisage such actions happening spontaneously at the quantum level, making the
small molecule itself into a quantum knot.
\item Let $\psi(A)$ be a function of a gauge field $A.$ Let
$$\hat{\psi}(K) = \int {\cal DA} \psi(A) {\cal H}_{K}(A),$$
where the integral denotes your favorite notion of integrating over gauge fields (one chooses a heuristic, or fixes the gauge to allow a measure theory that
can work) and ${\cal H}_{K}(A)$ denotes the trace of the holonomy of the gauge field taken around the specific embedding of the knot $K$ in three dimensional
space. This is the {\em loop transform } of the function $\psi(A)$ to a function $\hat{\psi}(K)$  of knotted loops in three dimensional space. The loop transform
is not necessarily invariant under topological moves, but this is sometimes the case. We would like to, at least at the formal level, formulate an inverse
transform to the loop transform. This would take the form 
$$\check{\phi}(A) = \sum_{K \in {\bf K}} \phi(K) {\cal H}_{K}(A) = \phi(\sum_{K \in {\bf K}} {\cal H}_{K}(A)) |K \rangle)$$
where $\phi(K)$ is a functional on knots and these sums would receive appropriate normalizations. Note that $\check{\phi}(A) = \phi ({\cal Q_{H}(A)})$
where ${\cal Q_{H}(A)}$ is the quantum knot
$${\cal Q_{H}(A)} = \sum_{K \in {\bf K}} {\cal H}_{K}(A)) |K \rangle.$$
While it is impractical to consider integrating over all
possible embeddings of a circle into three dimensional space, it is mathematically possible to examine summations involving all knot types. In this way the 
notion of quantum knot is inextricably tied to these questions about the loop transform. The loop transform is of particular value in the quantum gravity theory
of Ashtekar, Smolin and Rovelli \cite{SR}.
\item State summmation models for knot invariants such as the bracket state sum model \cite{Kspin,KP} for the Jones polynomial use collections of internal states
for a given knot diagram. Thus one has formulas such as 
$$\langle K \rangle = \sum_{S \in \cal S} \langle K | S \rangle $$
where, in the case of the bracket polynomial $\langle K | S \rangle$ is a product of vertex weights multiplied by a ``loop value" raised to the number of loops
in the state $S.$ The set of states $\cal S$ is obtained combinatorially from the diagram. (This description differs slightly in notation from that used in 
the references.) We see that it is natural to write
$$|K \rangle = \sum \langle K | S \rangle | S \rangle,$$
writing a quantum knot state in terms of its internal states. Then with
$|{\cal S}> = \sum |S \rangle,$ we have
$$\langle {\cal S}| K \rangle  = \sum \langle K | S \rangle = \langle K \rangle.$$
In this formalism, one can regard the state $|K \rangle = \sum \langle K | S \rangle | S \rangle$ as a preparation, and the computation
$\langle {\cal S}| K \rangle$ as the relative amplitude for measurement in the state $| {\cal S} \rangle.$ Thus this is a schema for quantum 
computation (albeit inefficient) of these invariants.
\end{enumerate}

\section{The Aravind Hypothesis}
In this section we consider analogies between knots and quantum states. This realm of analogies is a departure from the strict definition of quantum knots
of the previous sections, but is related to it in ways that deserve investigation.
A topological
entanglement is a non-local feature of a topological system. A quantum
entanglement is a non-local feature of a quantum system. Take the case
of the Hopf link of linking number one. See Figure 3.  In this Figure we show a
simple link of two components and state its inequivalence to the disjoint union
of two unlinked loops.  The analogy that one wishes to draw is with a state of
the form $$\psi = (|01 \rangle - |10 \rangle)/\sqrt{2}$$ which is {\em quantum entangled.}
That is, this state is not of the form $\psi_{1} \otimes \psi_{2} \in H \otimes H$ where $H$
is a complex vector space of dimension two. Cutting a component of the link removes
its topological entanglement. Observing the state removes its quantum entanglement in this case.
\bigbreak

\begin{center}
{\tt    \setlength{\unitlength}{0.92pt} \begin{picture}(282,127) \thicklines  
\put(240,45){\framebox(41,41){}} \put(191,45){\framebox(41,41){}}
\put(168,85){\line(-2,-3){21}} \put(166,69){\vector(-1,0){29}}
\put(150,69){\vector(1,0){24}} \put(78,84){\line(-1,0){34}}
\put(123,84){\line(-1,0){34}} \put(124,4){\line(0,1){80}}
\put(44,3){\line(1,0){80}} \put(44,84){\line(0,-1){81}}
\put(3,45){\line(0,1){78}} \put(36,44){\line(-1,0){32}}
\put(84,44){\line(-1,0){36}} \put(85,124){\line(0,-1){80}}
\put(3,124){\line(1,0){82}} \end{picture}}

{\bf Figure 3 - The Hopf Link} \bigbreak
\end{center}

Aravind \cite{Ara} proposed that
the entanglement of a link should correspond to 
the entanglement of a state. {\em Observation of a link 
would be  modeled by deleting one
component of the link.}
\bigbreak
 
In Figure 4 we illustrate the Borrommean rings. These rings are topologically linked, but any two of them,
taken alone, are unlinked. This description of the topological property of these rings can be called the {\it entanglement pattern} of the rings.
For example, it is not hard to design three rings with the linking pattern that all three are linked, and, if you remove any one of them then the other two are
also linked. The problem of classifying all links with a given linking pattern is an open problem in the theory of knots and links. A collection of $n$ rings is
said to be $\it Brunnian$ if the totality of the $n$ rings is linked, but upon removal of any one of them, the remaining rings are totally unlinked. There are 
many ways to design examples of Brunnian links. In Figure 5 we indicate one such method by illustrating the Borrommean rings as the closure of a braid $B$, and
then indicating a scheme for making a new braid from $B$ whose closure will be Brunnian (for $n = 4$). If we say that {\it a braid is Brunnian} if its closure
(obtained by attaching the bottom strands to the top strands as shown in Figure 5) is Brunnian, then this procedure produces a new Brunnian braid from any
given  Brunnian Braid. In this way one can design links with given patterns of entanglement.
\bigbreak

\begin{center}
{\tt    \setlength{\unitlength}{0.92pt} \begin{picture}(145,141) \thicklines  
\put(61,138){\line(1,0){81}} \put(142,23){\line(0,1){114}}
\put(122,22){\line(1,0){19}} \put(61,22){\line(1,0){50}}
\put(61,75){\line(0,-1){53}} \put(61,138){\line(0,-1){51}}
\put(116,83){\line(-1,0){11}} \put(116,3){\line(0,1){80}}
\put(3,3){\line(1,0){113}} \put(3,82){\line(0,-1){80}}
\put(13,82){\line(-1,0){10}} \put(28,82){\line(1,0){62}}
\put(21,42){\line(0,1){79}} \put(53,42){\line(-1,0){32}}
\put(99,42){\line(-1,0){34}} \put(102,42){\line(0,0){0}}
\put(101,122){\line(0,-1){80}} \put(69,122){\line(1,0){32}}
\put(21,122){\line(1,0){32}} \end{picture}}

{\bf Figure 4 -  Borrommean Rings}
\end{center}

\noindent 
Deleting any component of the Borommean rings 
yields a remaining pair of unlinked rings. The Borrommean 
rings are entangled, but any two of them are unentangled.
In this sense the Borrommean rings are analogous to 
the $GHZ$ state $|GHZ \rangle  = (1/\sqrt{2})(|000\rangle  + |111\rangle )$.
Observation in any factor of the $GHZ$ yields 
an unentangled state.  
\bigbreak

One can generalize the correspondence by taking, a state $$|GHZ[n] \rangle = (1/\sqrt{2})(|000 \cdots 0 \rangle + |111 \cdots 1 \rangle)$$
where there are $n$ tensor factors. This state has the same entanglement pattern as a Brunnian link of $n$ components.
\bigbreak

Aravind points out that 
{\em this correspondence property of quantum states and topological links is basis dependent.} To see this, we will calculate with unnormalized states.
Let $$|\psi \rangle = |000 \rangle + |111 \rangle.$$
Take a new basis $$\{ |0' \rangle, |1' \rangle \}$$ where
$$| 0 \rangle = | 0' \rangle + | 1' \rangle$$
$$| 1 \rangle = | 0' \rangle - | 1' \rangle.$$
Then
$$|\psi \rangle = |0'00 \rangle + |1'00 \rangle + |0'11 \rangle - |1'11 \rangle,$$
and we use the new basis in the first tensor factor, and the old basis in the other two tensor factors.
Thus
$$|\psi \rangle = |0' \rangle [|00 \rangle + |11 \rangle] + |1' \rangle [|00 \rangle - |11 \rangle],$$
$$|\psi \rangle = [|0 \rangle]_{2} [|0'0 \rangle + |1'0 \rangle]_{13} + [|1 \rangle]_{2} [|0'1 \rangle - |1'1 \rangle]_{13},$$
$$|\psi \rangle = [|0'0 \rangle + |1'0 \rangle] |0 \rangle  +  [|0'1 \rangle - |1'1 \rangle]|1 \rangle.$$
With this change of basis, observation in the first factor yields an entangled state, but observation in either the second or the 
third factors yield unentangled states! If we wanted a link that had these entanglement properties (for cutting components) we could
choose the link $L$ shown in Figure 6. This link has three components labeled $1$, $2$ and $3.$ Removal of component number $1$ leaves 
the topologically non-trivial Hopf link. Removal of either component $2$ or component $3$ leaves two unlinked curves. Thus this link
$L$ is analogous to the second version of the $GHZ$ state.
\bigbreak

\begin{center}
{\tt    \setlength{\unitlength}{0.92pt}
\begin{picture}(283,332)
\thinlines    \put(207,125){$B^{-1}$}
\thicklines   \put(53,251){\line(-1,1){15}}
              \put(13,291){\line(1,-1){15}}
              \put(53,291){\line(-1,-1){41}}
              \put(52,211){\line(1,1){17}}
              \put(92,251){\line(-1,-1){16}}
              \put(52,251){\line(1,-1){41}}
              \put(13,250){\line(0,-1){39}}
              \put(92,291){\line(0,-1){40}}
              \put(93,209){\line(0,-1){40}}
              \put(13,169){\line(0,-1){39}}
              \put(52,170){\line(1,-1){41}}
              \put(92,170){\line(-1,-1){16}}
              \put(52,130){\line(1,1){17}}
              \put(53,211){\line(-1,-1){41}}
              \put(13,210){\line(1,-1){15}}
              \put(53,170){\line(-1,1){15}}
              \put(93,129){\line(0,-1){40}}
              \put(12,88){\line(0,-1){39}}
              \put(50,92){\line(1,-1){41}}
              \put(91,89){\line(-1,-1){16}}
              \put(51,49){\line(1,1){17}}
              \put(52,129){\line(-1,-1){41}}
              \put(12,129){\line(1,-1){15}}
              \put(52,89){\line(-1,1){15}}
              \put(93,291){\line(1,0){20}}
              \put(114,290){\line(0,-1){240}}
              \put(114,51){\line(-1,0){22}}
              \put(53,291){\line(0,1){12}}
              \put(53,303){\line(1,0){80}}
              \put(133,302){\line(0,-1){269}}
              \put(133,33){\line(-1,0){81}}
              \put(52,32){\line(0,1){18}}
              \put(13,291){\line(0,1){29}}
              \put(14,320){\line(1,0){137}}
              \put(150,321){\line(0,-1){309}}
              \put(150,12){\line(-1,0){138}}
              \put(12,11){\line(0,1){40}}
              \put(189,226){\framebox(42,40){}}
              \put(230,185){\line(1,1){17}}
              \put(270,225){\line(-1,-1){16}}
              \put(230,225){\line(1,-1){41}}
              \put(230,146){\line(1,1){17}}
              \put(271,183){\line(-1,-1){16}}
              \put(230,187){\line(1,-1){41}}
              \put(191,105){\framebox(40,41){}}
              \put(270,66){\line(-1,1){15}}
              \put(230,106){\line(1,-1){15}}
              \put(271,106){\line(-1,-1){41}}
              \put(270,25){\line(-1,1){15}}
              \put(230,65){\line(1,-1){15}}
              \put(270,65){\line(-1,-1){41}}
              \put(190,227){\line(0,-1){82}}
              \put(191,106){\line(0,-1){80}}
              \put(269,309){\line(0,-1){83}}
              \put(229,307){\line(0,-1){41}}
              \put(189,306){\line(0,-1){42}}
              \put(271,145){\line(0,-1){40}}
              \put(206,245){$B$}
\end{picture}}

{\bf Figure 5 -  Borrommean Braid and Template}
\end{center}

\begin{center}
{\tt    \setlength{\unitlength}{0.92pt}
\begin{picture}(267,214)
\thicklines   \put(236,85){$3$}
              \put(196,192){$2$}
              \put(38,45){$1$}
              \put(209,106){\line(0,-1){94}}
              \put(129,154){\line(0,-1){32}}
              \put(129,170){\line(0,-1){8}}
              \put(128,201){\line(0,-1){24}}
              \put(12,12){\line(1,0){197}}
              \put(12,189){\line(0,-1){176}}
              \put(121,189){\line(-1,0){108}}
              \put(156,191){\line(-1,0){20}}
              \put(157,175){\line(0,1){16}}
              \put(34,174){\line(1,0){123}}
              \put(34,38){\line(0,1){136}}
              \put(188,37){\line(-1,0){154}}
              \put(159,158){\line(0,-1){18}}
              \put(66,158){\line(1,0){93}}
              \put(66,89){\line(0,1){67}}
              \put(134,139){\line(1,0){23}}
              \put(84,140){\line(1,0){39}}
              \put(84,107){\line(0,1){33}}
              \put(250,81){\line(-1,0){35}}
              \put(193,80){\line(1,0){12}}
              \put(176,106){\line(1,0){33}}
              \put(170,80){\line(1,0){14}}
              \put(188,88){\line(0,-1){49}}
              \put(176,89){\line(1,0){11}}
              \put(67,89){\line(1,0){97}}
              \put(83,107){\line(1,0){82}}
              \put(203,162){\line(-1,0){34}}
              \put(248,162){\line(-1,0){34}}
              \put(249,81){\line(0,1){80}}
              \put(169,162){\line(0,-1){81}}
              \put(161,122){\line(-1,0){32}}
              \put(209,122){\line(-1,0){36}}
              \put(210,202){\line(0,-1){80}}
              \put(128,202){\line(1,0){82}}
\end{picture}}

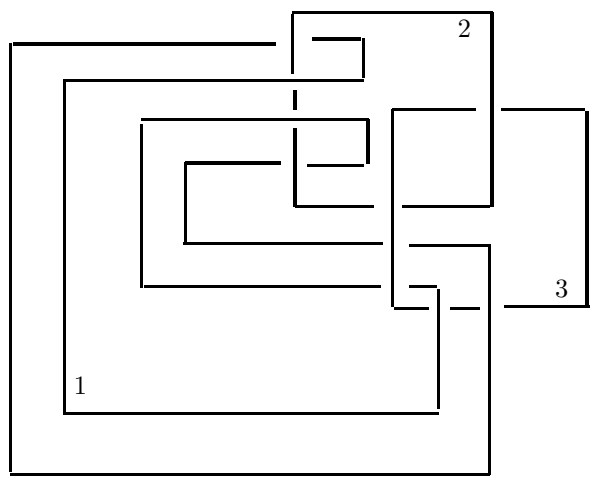
{\bf Figure 6 -  A Second Link $L$ for the $GHZ$ State}
\end{center}

If one follows the Aravind Hypothesis, there will be a multiplicity of links and 
entanglement patterns that correspond to a single quantum state. This sort of multiplicity leads to the notion that a
link might be considered as a kind of ``classical property" of a quantum state. In the work of Rasetti and Regge on quantum vortices \cite{RR}it is suggested
just this: that a knot or link could be regarded as the consequence of observing a quantum state of super-cooled helium, just as an eigenvalue is regarded as
the consequence of observing the state of an atom. We bring up their work to point out that the notion of a ``quantum knot" has existed in the physics
literature for some time. In the case of the work of Rasetti and Regge, the details of the classical knot corresponding to the quantum vortex
are extracted by a collection of operators that are applied to the quantum state. It is quite possible that there will
also be a multiplicity of classical knots associated with a given quantum circumstance. In the examples we have shown for the 
Aravind Hypothesis there is not enough physical substance to the quantum side of the picture to single out any given knot or link, or
even a collection of knots and links that would correspond to the quantum states. Nevertheless, the Aravind idea can be regarded as 
an abstraction of the more physical context of quantum knots in the sense of Regge and Rasetti. Quantum knots in this physical sense are
to be regarded as the results of an experimentalist attempting to elucidate the embedding geometry/topology of a vortexing phenomenon
that occurs on such a small scale that it cannot be seen directly in a classical manner. The resulting knots are then descriptions of
some aspects of the quantum state and possibly dependent upon choices of measurement apparatus.
\bigbreak

Another sort of quantum knot has been discussed by Sir Michael Berry and his collaborators \cite{Berry}. Berry's knot is the set of zeros
of a wave function defined on three dimensional space. The set can contain knotted curves, and Berry shows that this is indeed the case
for certain states of the hydrogen atom. Such quantum knots are the exact opposite of the Rasetti-Regge quantum knots. Berry's knotted
zeros are the places where nothing can be observed! They are the loci of destructive interference, not the loci of vortex action.
Clearly more work needs to be done in understanding quantum knotting at this physical level.
\bigbreak

\subsection{Quantum Entanglement and Probabilistic Knots}
Continuing the Aravind Analogy, we now point out 
that {\em there are quantum states whose entanglement 
after an observation is a matter of probability} (via computation of quantum amplitudes).
\bigbreak

\noindent Consider the state 

{\bf \[
|\psi \rangle=(1/2)(|000\rangle  + |001\rangle  + |101\rangle  + |110\rangle ).
\]} 

\noindent Observation in any coordinate yields an entangled or an
unentangled state with equal probability. For example

{\bf \[
|\psi \rangle=(1/2)(|0\rangle(|00\rangle  + |01\rangle)  + |1\rangle(|01\rangle  + |10\rangle )
\] }

\noindent so that projecting to $|0\rangle$ in the first coordinate yields an unentangled state, while projecting to $|1\rangle$ yields an
entangled state, each with equal probability.
\bigbreak

If we wish to have a link,$B'$, analogous to the Borrommean rings, that models this state, we will need something new. The result of
cutting a component of $B'$ will have to yield up either a linked link or an unlinked link with probability $1/2$ for each.
One can imagine a mechanical scenario for this, as illustrated in Figure 7. In that Figure we show a copy of the Borrommean rings with 
extra influences of each component on one of the crossings in the link. When a component is cut, this extra influence causes the
corresponding crossing to switch with probability $1/2.$ Should we say that the state $| \psi \rangle$ above corresponds, by Aravind
Hypothesis, to the probabilistic link of Figure 7?  If we follow this line, then there will be a complexity of matching probability
amplitudes for quantum states with essentially classical probabilities for a class of links with extra structure.
\bigbreak

\begin{center}
{\tt    \setlength{\unitlength}{0.92pt}
\begin{picture}(164,186)
\thinlines    \put(14,13){A Probabilistic Link}
              \put(30,158){\vector(1,-1){31}}
              \put(149,172){\vector(-2,-3){30}}
              \put(12,38){\vector(3,2){49}}
              \put(71,78){\circle{22}}
              \put(110,117){\circle{22}}
              \put(71,116){\circle{22}}
\thicklines   \put(70,174){\line(1,0){81}}
              \put(151,59){\line(0,1){114}}
              \put(131,58){\line(1,0){19}}
              \put(70,58){\line(1,0){50}}
              \put(70,111){\line(0,-1){53}}
              \put(70,174){\line(0,-1){51}}
              \put(125,119){\line(-1,0){11}}
              \put(125,39){\line(0,1){80}}
              \put(12,39){\line(1,0){113}}
              \put(12,118){\line(0,-1){80}}
              \put(22,118){\line(-1,0){10}}
              \put(37,118){\line(1,0){62}}
              \put(30,78){\line(0,1){79}}
              \put(62,78){\line(-1,0){32}}
              \put(108,78){\line(-1,0){34}}
              \put(110,158){\line(0,-1){80}}
              \put(78,158){\line(1,0){32}}
              \put(30,158){\line(1,0){32}}
\end{picture}}

{\bf Figure 7 -  A Probabilistic Link}
\end{center}

New ways to use link diagrams must be invented to map the properties 
of such states. {\em We take
seriously the problem of classifying the 
entanglement patterns of quantum states.} We are convinced 
that such a classification will be of
practical importance to quantum computing, and quantum information theory.
\bigbreak

\section{Diagrammatic Methods for Quantum Measurement}
The point of view of this paper is based on diagrammatic conventions for matrix multiplication and tensor composition.
The purpose of this section is to describe these conventions and to show how they are used in our work. We take diagrams to represent matrices and 
products or concatenations of matrices. In this way a complex network diagram can represent a contraction of a collection of multi-indexed matrices, and so may 
represent a quantum state or a quantum amplitude. We regard each graph as both a possible holder for matrices, and hence as a vehicle for such a computation, 
{\em and} 
as a combinatorial structure. As a combinatorial structure the graph can be modified. An edge can be removed. A node can be inserted. Such modifications
can be interpreted in terms of quantum preparation and measurement. One can then take the graph as a miniature ``world" upon which such operations are performed.
Since the graphs can also represent topological structures, this approach leads to a way to interface topology with the quantum mechanics.
The diagrammatics in this section should be compared with work of Roger Penrose \cite{Pen}, the first author \cite{KP,Kspin} and Tom Etter \cite{Etter}.
\bigbreak

First, consider the multiplication of matrices $M = (M_{ij})$ and $N=(N_{kl})$ where $M$ is $m \times n$ and $N$ is $n \times p.$
Then $MN$ is $m \times p$ and $$(MN)_{ij} = \Sigma_{k=1}^{n} M_{ik}N_{kj}.$$  We represent each matrix by a box, and each index for 
the matrix elements by a line segment that is attached to this box. The common index in the summation is represented by a line that 
emanates from one box, and terminates in the other box. This line segment has no free ends. By the definition of matrix multipliction,
a line segment without free ends represents the summation over all possible index assignments that are available for that segment.
Segments with free ends correspond to the possible index choices for the product matrix. See Figure 8.
\bigbreak

\begin{center}
{\tt    \setlength{\unitlength}{0.92pt}
\begin{picture}(353,116)
\thinlines    \put(49,65){\framebox(41,41){}}
              \put(169,65){\framebox(41,40){}}
              \put(50,13){\framebox(41,41){}}
              \put(130,13){\framebox(41,41){}}
              \put(59,82){$M$}
              \put(180,82){$N$}
              \put(60,32){$M$}
              \put(140,32){$N$}
              \put(68,84){\line(0,0){0}}
              \put(130,32){\line(-1,0){40}}
              \put(50,33){\line(-1,0){40}}
              \put(170,33){\line(1,0){40}}
              \put(89,85){\line(1,0){22}}
              \put(146,86){\line(1,0){23}}
              \put(209,85){\line(1,0){21}}
              \put(49,86){\line(-1,0){20}}
              \put(229,29){$=$}
              \put(286,10){\framebox(40,42){}}
              \put(285,32){\line(-1,0){21}}
              \put(325,31){\line(1,0){18}}
              \put(292,28){$MN$}
\end{picture}}

{\bf Figure 8 -  Matrix Multiplication}
\end{center}

\noindent The trace of an $m \times m$ matrix $M$ is given by the formula $$tr(M) = \Sigma_{i=1}^{m} M_{ii}.$$
In diagrammatic terms the trace is represented by a box with the output segment identified with the input segment.
See Figure 9 for this interpretation of the matrix trace. 
\bigbreak

\begin{center}
{\tt    \setlength{\unitlength}{0.92pt}
\begin{picture}(186,92)
\thinlines    \put(113,48){$= tr(M)$}
              \put(53,29){$M$}
              \put(12,35){\line(0,1){45}}
              \put(12,35){\line(1,0){26}}
              \put(107,80){\line(-1,0){95}}
              \put(107,34){\line(0,1){46}}
              \put(90,34){\line(1,0){17}}
              \put(38,10){\framebox(53,44){}}
\end{picture}}

{\bf Figure 9 -  Matrix Trace}
\end{center}

\begin{center}
{\tt    \setlength{\unitlength}{0.92pt}
\begin{picture}(425,279)
\thinlines    \put(61,165){$tr(M)$}
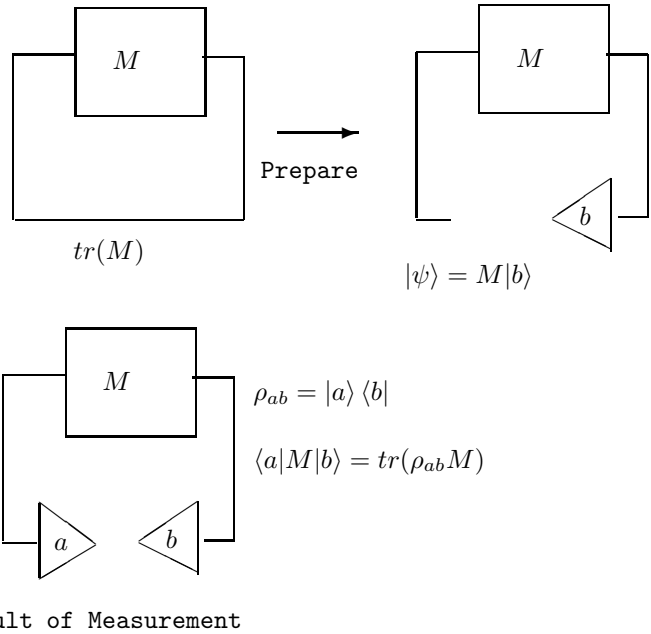
\thicklines   \put(10,13){Result of Measurement}
              \put(138,198){Prepare}
              \put(145,217){\vector(1,0){33}}
\thinlines    \put(122,181){\line(-1,0){85}}
              \put(131,248){\line(0,-1){67}}
              \put(36,249){\line(0,-1){68}}
              \put(62,224){\framebox(53,44){}}
              \put(114,248){\line(1,0){17}}
              \put(36,249){\line(1,0){26}}
              \put(77,243){$M$}
              \put(131,181){\line(-1,0){12}}
              \put(197,155){$|\psi \rangle = M| b \rangle$}
              \put(297,182){\line(-1,0){12}}
              \put(202,181){\line(1,0){14}}
              \put(243,244){$M$}
              \put(202,250){\line(1,0){26}}
              \put(280,249){\line(1,0){17}}
              \put(228,225){\framebox(53,44){}}
              \put(202,250){\line(0,-1){68}}
              \put(297,249){\line(0,-1){67}}
              \put(282,198){\line(0,-1){30}}
              \put(282,198){\line(-3,-2){24}}
              \put(258,181){\line(2,-1){24}}
              \put(269,179){$b$}
              \put(99,46){$b$}
              \put(53,45){$a$}
              \put(88,48){\line(2,-1){24}}
              \put(112,65){\line(-3,-2){24}}
              \put(112,65){\line(0,-1){30}}
              \put(70,47){\line(-5,-3){23}}
              \put(47,65){\line(4,-3){23}}
              \put(47,64){\line(0,-1){30}}
              \put(127,116){\line(0,-1){67}}
              \put(32,117){\line(0,-1){68}}
              \put(58,92){\framebox(53,44){}}
              \put(110,116){\line(1,0){17}}
              \put(32,117){\line(1,0){26}}
              \put(73,111){$M$}
              \put(32,48){\line(1,0){14}}
              \put(127,49){\line(-1,0){12}}
              \put(135,107){$\rho_{ab} = | a \rangle \, \langle b |$}
              \put(135,79){$\langle a | M | b \rangle = tr(\rho_{ab} M)$}
\end{picture}}

{\bf Figure 10 -  Network Operations: Preparation and Measurement via Insertion of Bras and Kets.}
\end{center}

In Figure 10 we illustrate the diagrammatic interpretation of the formula
$$\langle a | M | b \rangle = tr(\rho_{ab} M).$$
This formula gives the amplitude for measuring the state $|a \rangle$ from a preparation of $|\psi \rangle = M | b \rangle.$
Note that the state $| \psi \rangle$ is obtained from the graphical structure of $tr(M)$ by cutting the connection between the 
input line and output line of the box labeled $M,$ and inserting the ket $| b \rangle$ on the output line. The resulting network
is shown at the top of Figure 10. This network, with one free end, represents the quantum state $| \psi \rangle.$ This state is the superposition
of all possible values (qubits) that can occur at the free end of the network. When we measure the state, one of the possible qubits occurs. 
The amplitude for the occurrence of $| a \rangle$ is equal to $\langle a | M | b \rangle.$ When we insert $| a \rangle$ at the free end of the network
for $| \psi \rangle,$ we obtain the network whose value is this amplitude.
\bigbreak 
 
\noindent If $M$ is unitary, then we can interpret the formula as the amplitude for measuring state $|a \rangle$ from 
a preparation in $| b \rangle,$ and an evolution of this preparation by the unitary transformation $M.$ In the first
interpretation the operator $M$ can be an observable, aiding in the preparation of the state. In this notation,
$$\rho_{ab} = | a \rangle \, \langle \, b \, |$$ is the ket-bra associated with the states $| a \rangle$ and $| b \rangle.$
If $a=b$, then $\rho_{aa}$ is the density matrix associated with the pure state $| a \rangle.$ 
\bigbreak

The key to this graphical model for preparation and measurement is the understanding that the diagram is both a combinatorial structure {\it and} a representative
for the computation of either an amplitude or a state (via summation over the indices available for the internal lines and superposition over the possibilities for
the free ends of the network). A diagram with free ends (no kets or bras tied into the ends) represents a state that is the superposition of all the possibilities
for the values of the free ends. This superposition is a superposition of diagrams with different labels on the ends. In this way the principles of quantum
measurement are seen to live in categories of diagrams. A given diagram can be regarded as a world that is subject to preparation and measurement. After such
an operation is performed (Cut an edge. Insert a density matrix.), a new world is formed that is itself subject to preparation and measurement. This succession 
of worlds and states can be regarded as a description of the evolution of a quantum process. 
\bigbreak

\noindent {\bf Remark.} One can generalize this notion of quantum process in networks by allowing the insertion of other operators into the network, and by
allowing systematic operations on the graph. Techniques of this sort are used in spin foam models for quantum gravity \cite{M}, and in renormalization of
statistical mechanics models.
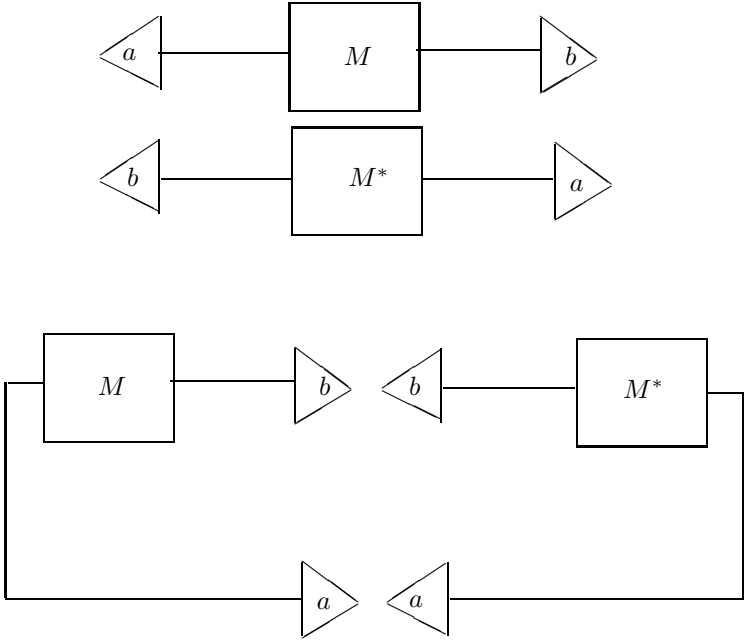
\bigbreak

\begin{center}
{\tt    \setlength{\unitlength}{0.92pt}
\begin{picture}(482,319)
\thinlines    \put(10,13){$|\langle a | M | b \rangle |^{2} = tr(\rho_{bb} M^{*} \rho_{aa} M)$}
              \put(175,234){$M^{*}$}
              \put(203,290){\line(1,0){51}}
              \put(97,289){\line(1,0){54}}
              \put(152,214){\framebox(53,44){}}
              \put(254,303){\line(0,-1){30}}
              \put(254,304){\line(4,-3){23}}
              \put(277,286){\line(-5,-3){23}}
              \put(82,286){$a$}
              \put(97,253){\line(0,-1){30}}
              \put(97,253){\line(-3,-2){24}}
              \put(73,236){\line(2,-1){24}}
              \put(84,234){$b$}
              \put(264,284){$b$}
              \put(73,287){\line(2,-1){24}}
              \put(97,304){\line(-3,-2){24}}
              \put(98,304){\line(0,-1){30}}
              \put(266,232){$a$}
              \put(283,234){\line(-5,-3){23}}
              \put(260,252){\line(4,-3){23}}
              \put(260,251){\line(0,-1){30}}
              \put(173,284){$M$}
              \put(151,265){\framebox(53,44){}}
              \put(98,237){\line(1,0){54}}
              \put(205,237){\line(1,0){54}}
              \put(288,147){$M^{*}$}
              \put(214,151){\line(1,0){54}}
              \put(50,129){\framebox(53,44){}}
              \put(72,148){$M$}
              \put(156,79){\line(0,-1){30}}
              \put(156,80){\line(4,-3){23}}
              \put(179,62){\line(-5,-3){23}}
              \put(162,60){$a$}
              \put(216,79){\line(0,-1){30}}
              \put(215,79){\line(-3,-2){24}}
              \put(191,62){\line(2,-1){24}}
              \put(163,148){$b$}
              \put(200,148){$b$}
              \put(189,150){\line(2,-1){24}}
              \put(213,167){\line(-3,-2){24}}
              \put(213,167){\line(0,-1){30}}
              \put(200,61){$a$}
              \put(176,150){\line(-5,-3){23}}
              \put(153,168){\line(4,-3){23}}
              \put(153,167){\line(0,-1){30}}
              \put(269,127){\framebox(53,44){}}
              \put(102,154){\line(1,0){51}}
              \put(50,153){\line(-1,0){15}}
              \put(337,149){\line(-1,0){15}}
              \put(34,153){\line(0,-1){88}}
              \put(337,149){\line(0,-1){85}}
              \put(34,64){\line(1,0){121}}
              \put(337,64){\line(-1,0){120}}
\end{picture}}

{\bf Figure 11 - Probability by Mating a Network with its Dual Network}
\end{center}

In Figure 11 we illustrate a diagrammatic interpretation of the formula
$$|\langle a | M | b \rangle |^{2} = tr(\rho_{bb} M^{*} \rho_{aa} M),$$
expressing the probability corresponding to the probability amplitude
$\langle a | M | b \rangle.$ If $\cal N$ is the network corresponding to $tr(M)$, and $\cal N'$ is the network
corresponding to cutting $\cal N$ and inserting the bra and the ket, then the network for $|\langle a | M | b \rangle |^{2}$
can be described as the {\em double}, $$D(\cal N) = \cal{N} \cal{N^*}.$$ Here $\cal N^{*}$ is obtained from $\cal N$ by taking the transposed conjugate $M^{*}$ of
$M,$ cutting the $tr(M^{*})$ network and inserting the ket and bra. The networks $\cal N$ and $\cal N^{*}$ are then juxtaposed so that 
density matrices appear at the juxtapositions, and we get the diagram for the formula above, for the probability amplitude.
\bigbreak

\begin{center}
{\tt    \setlength{\unitlength}{0.92pt}
\begin{picture}(347,570)
\thicklines   \put(207,25){\vector(0,1){50}}
              \put(101,13){Note insertion of density matrix.}
\thinlines    \put(133,213){\line(3,-5){29}}
              \put(129,485){\line(2,-3){33}}
              \put(335,166){\line(-1,-4){6}}
              \put(307,158){\line(4,1){27}}
              \put(308,157){\line(4,-3){20}}
              \put(288,112){\line(5,6){30}}
              \put(223,191){\line(5,-6){65}}
              \put(27,44){\line(1,1){14}}
              \put(27,45){\line(-1,2){7}}
              \put(21,59){\line(1,0){19}}
              \put(120,32){\line(-1,3){7}}
              \put(100,30){\line(1,0){20}}
              \put(113,53){\line(-3,-5){14}}
              \put(131,36){\line(0,1){25}}
              \put(148,50){\line(-5,-4){17}}
              \put(127,61){\line(2,-1){21}}
              \put(235,81){\line(-4,-1){16}}
              \put(224,97){\line(3,-4){11}}
              \put(224,98){\line(-1,-4){5}}
              \put(304,214){\line(-3,-4){15}}
              \put(278,222){\line(4,-1){26}}
              \put(278,222){\line(2,-5){11}}
              \put(220,192){\line(4,1){64}}
              \put(186,248){\line(-1,0){15}}
              \put(182,234){\line(1,3){4}}
              \put(171,248){\line(4,-5){11}}
              \put(136,287){\line(0,-1){11}}
              \put(123,283){\line(4,1){13}}
              \put(124,283){\line(5,-3){13}}
              \put(157,101){\circle*{10}}
              \put(37,79){\circle*{10}}
              \put(47,80){\circle*{10}}
              \put(68,90){\circle*{10}}
              \put(119,107){\circle*{10}}
              \put(149,114){\circle*{10}}
              \put(144,126){\circle*{10}}
              \put(132,152){\circle*{10}}
              \put(85,137){\circle*{10}}
              \put(89,164){\circle*{10}}
              \put(93,175){\circle*{10}}
              \put(119,175){\circle*{10}}
              \put(163,163){\circle*{10}}
              \put(157,173){\circle*{10}}
              \put(148,193){\circle*{10}}
              \put(133,213){\circle*{10}}
              \put(120,207){\circle*{10}}
              \put(94,192){\circle*{10}}
              \put(63,181){\circle*{10}}
              \put(41,195){\circle*{10}}
              \put(18,209){\circle*{10}}
              \put(90,219){\circle*{10}}
              \put(83,250){\circle*{10}}
              \put(95,251){\circle*{10}}
              \put(223,192){\circle*{10}}
              \put(174,193){\circle*{10}}
              \put(160,229){\circle*{10}}
              \put(290,113){\circle*{10}}
              \put(113,251){\circle*{10}}
              \put(111,251){\line(2,3){18}}
              \put(47,80){\line(3,-2){57}}
              \put(158,231){\line(2,-5){14}}
              \put(155,101){\line(5,-1){67}}
              \put(156,101){\line(-2,-5){18}}
              \put(94,251){\line(3,-5){26}}
              \put(146,193){\line(1,0){76}}
              \put(129,151){\line(5,2){33}}
              \put(83,138){\line(-1,-3){16}}
              \put(84,137){\line(6,-5){35}}
              \put(35,78){\line(3,1){114}}
              \put(62,182){\line(-1,-4){31}}
              \put(39,195){\line(2,1){48}}
              \put(118,175){\line(1,-2){37}}
              \put(157,175){\line(-1,0){65}}
              \put(82,251){\line(1,0){48}}
              \put(18,208){\line(5,-3){72}}
              \put(94,192){\line(-1,-5){11}}
              \put(95,192){\line(5,3){81}}
              \put(83,251){\line(1,-5){12}}
              \put(18,208){\line(3,2){64}}
              \put(17,211){\line(1,2){16}}
              \put(17,252){\line(2,-1){29}}
              \put(16,252){\line(5,2){27}}
              \put(43,264){\line(1,-5){6}}
              \put(130,262){\line(0,-1){24}}
              \put(130,262){\line(2,-1){21}}
              \put(151,250){\line(-2,-1){21}}
              \put(148,522){\line(-2,-1){21}}
              \put(127,534){\line(2,-1){21}}
              \put(127,534){\line(0,-1){24}}
              \put(40,536){\line(1,-5){6}}
              \put(13,524){\line(5,2){27}}
              \put(14,524){\line(2,-1){29}}
              \put(14,483){\line(1,2){16}}
              \put(183,130){\line(0,-1){30}}
              \put(183,131){\line(4,-3){23}}
              \put(207,113){\line(-5,-3){23}}
              \put(259,129){\line(0,-1){30}}
              \put(259,129){\line(-3,-2){24}}
              \put(235,112){\line(2,-1){24}}
              \put(189,111){$a$}
              \put(246,110){$b$}
              \put(15,480){\line(3,2){64}}
              \put(80,523){\line(1,-5){12}}
              \put(92,464){\line(5,3){81}}
              \put(91,464){\line(-1,-5){11}}
              \put(15,480){\line(5,-3){72}}
              \put(79,523){\line(1,0){48}}
              \put(154,447){\line(-1,0){65}}
              \put(115,447){\line(1,-2){37}}
              \put(36,467){\line(2,1){48}}
              \put(59,454){\line(-1,-4){31}}
              \put(32,350){\line(3,1){114}}
              \put(81,409){\line(6,-5){35}}
              \put(80,410){\line(-1,-3){16}}
              \put(126,423){\line(5,2){33}}
              \put(143,465){\line(1,0){76}}
              \put(91,523){\line(3,-5){26}}
              \put(153,373){\line(-2,-5){18}}
              \put(152,373){\line(5,-1){67}}
              \put(155,503){\line(2,-5){14}}
              \put(44,352){\line(3,-2){57}}
              \put(108,523){\line(2,3){18}}
              \put(110,523){\circle*{10}}
              \put(287,385){\circle*{10}}
              \put(157,501){\circle*{10}}
              \put(171,465){\circle*{10}}
              \put(220,464){\circle*{10}}
              \put(92,523){\circle*{10}}
              \put(80,522){\circle*{10}}
              \put(87,491){\circle*{10}}
              \put(15,481){\circle*{10}}
              \put(38,467){\circle*{10}}
              \put(60,453){\circle*{10}}
              \put(91,464){\circle*{10}}
              \put(117,479){\circle*{10}}
              \put(130,485){\circle*{10}}
              \put(145,465){\circle*{10}}
              \put(154,445){\circle*{10}}
              \put(160,435){\circle*{10}}
              \put(116,447){\circle*{10}}
              \put(90,447){\circle*{10}}
              \put(86,436){\circle*{10}}
              \put(82,409){\circle*{10}}
              \put(129,424){\circle*{10}}
              \put(141,398){\circle*{10}}
              \put(146,386){\circle*{10}}
              \put(116,379){\circle*{10}}
              \put(65,362){\circle*{10}}
              \put(44,352){\circle*{10}}
              \put(34,351){\circle*{10}}
              \put(154,373){\circle*{10}}
              \put(121,555){\line(5,-3){13}}
              \put(120,555){\line(4,1){13}}
              \put(133,559){\line(0,-1){11}}
              \put(168,520){\line(4,-5){11}}
              \put(179,506){\line(1,3){4}}
              \put(183,520){\line(-1,0){15}}
              \put(217,464){\line(4,1){64}}
              \put(275,494){\line(2,-5){11}}
              \put(275,494){\line(4,-1){26}}
              \put(301,486){\line(-3,-4){15}}
              \put(221,370){\line(-1,-4){5}}
              \put(221,369){\line(3,-4){11}}
              \put(232,353){\line(-4,-1){16}}
              \put(124,333){\line(2,-1){21}}
              \put(145,322){\line(-5,-4){17}}
              \put(128,308){\line(0,1){25}}
              \put(110,325){\line(-3,-5){14}}
              \put(97,302){\line(1,0){20}}
              \put(117,304){\line(-1,3){7}}
              \put(18,331){\line(1,0){19}}
              \put(24,317){\line(-1,2){7}}
              \put(24,316){\line(1,1){14}}
              \put(148,386){\line(1,0){139}}
              \put(220,463){\line(5,-6){65}}
              \put(285,384){\line(5,6){30}}
              \put(305,429){\line(4,-3){20}}
              \put(304,430){\line(4,1){27}}
              \put(332,438){\line(-1,-4){6}}
              \put(152,114){\line(1,0){29}}
              \put(258,113){\line(1,0){29}}
\thicklines   \put(293,355){\vector(0,-1){113}}
              \put(162,300){Prepare and Measure}
\end{picture}}

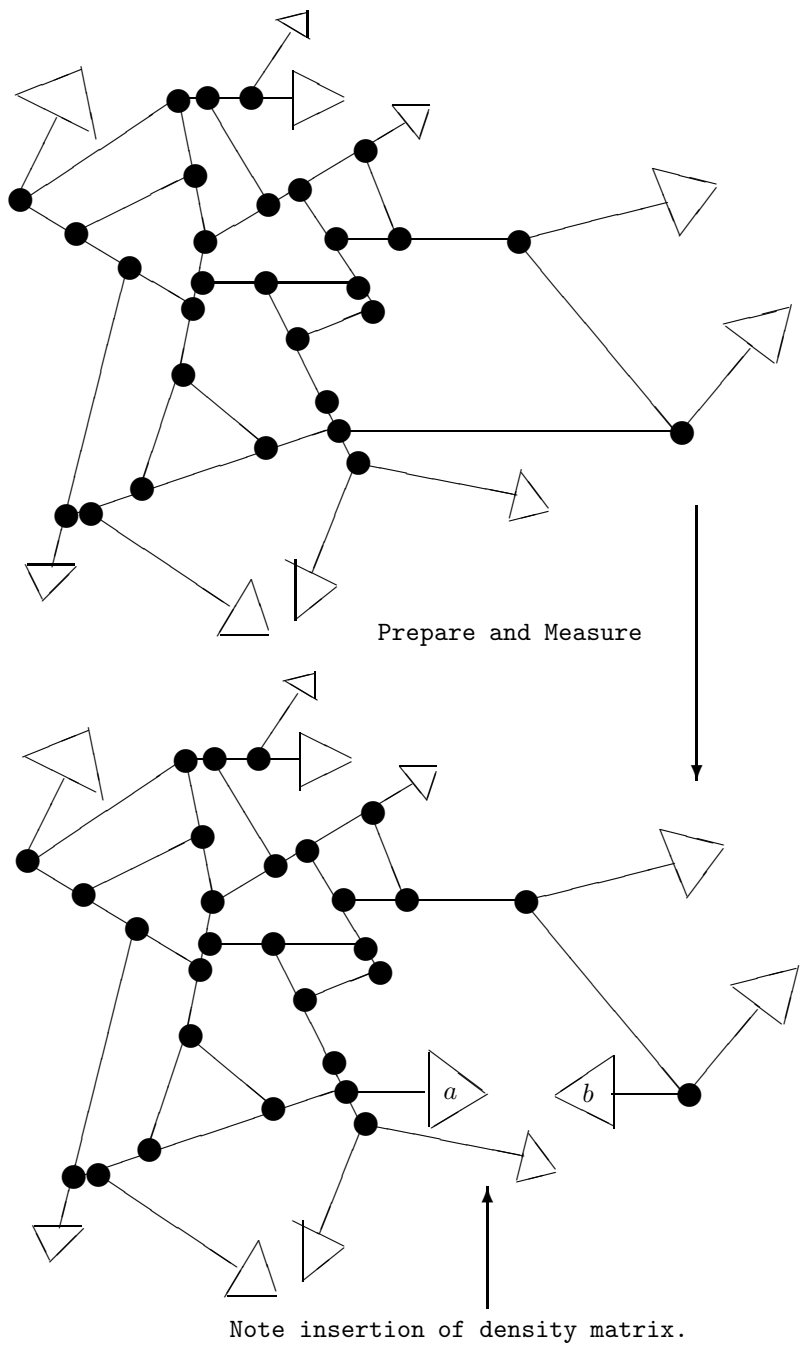
{\bf Figure 12 -  Preparing and Measuring in a General Network}
\end{center}

\noindent {\bf Generalizing to a Network.} 	In this diagrammatic interpretation, we obtain the amplitude from the closed loop diagram for
$tr(M)$ by cutting a segment from that diagram and inserting the ket $| a\, \rangle$ and the bra $| b \, \rangle.$
We can generalize this notion by thinking of the diagram for $tr(M)$ as a network (quantum network) wherein we have performed
a preparation and measurement by the operation of cutting an edge, and inserting a ket and a bra into the site of that edge.
Figure 12 illustrates exactly this idea with a sample trivalent network. Each node of the network corresponds to a matrix whose entries are determined by
assigments of labels to the edges incident to the node. The free ends of the network are decorated with kets to emphasize that the network has had specific state
choices at its free ends.  If some ends are left free, then the network represents a quantum state, as described above. One should think of the network as
representing the sum, over all assignments of states to its internal lines, of the products of matrix elements generated at the vertices of the  network.
Each specific network without free ends represents a quantum amplitude that is computed in this way. Each network is its own path integral. 
Figure 12 illustrates an act of preparation and measurement on the network. An edge is cut from the net, and a ket-bra is inserted at that edge. The processes
of cutting and insertion are local, but the resulting path sum computation (integral to the definition of the net) is changed in a global way. Note that the
insertion of a bra {\it and } a ket in the edge corresponds to one possible measurement outcome. The cutting of an edge with the insertion of a ket and an open
end (the result of the cut) represents the  quantum state so prepared, before any measurement has happened.
\bigbreak

If one imagines replacing 
the familiar Euclidean or differential geometric background space of quantum physics, with a network of this sort, then the
non-locality  of quantum mechanics is extended to the non-locality inherent in the network, and topological properties of the network
will have an interplay with this non-locality. We shall return to this theme after more discussion about diagrammatic representations.
\bigbreak  

Networks can be topological. In a paper on spin networks \cite{Kspin} by the first author there is an account of some of the relationships between knot theory
and a generalization of Penrose spin networks. It is not the purpose of this paper to go into great detail on this theory, but Penrose \cite{Pen}
originally designed his spin networks as a replacement for a background space, and he discovered that networks whose amplitudes
were invariant under successive observations had the property that they did indeed model directions in three dimensional space.
Later \cite{KA87,KA89,KP,KL,KaufInter} a generalization of the Penrose spin networks was found to encompass invariants of knots and links. 
Spin network structures can be used directly in quantum computing \cite{FLZ,MR,BG}.
In the generalization, each knot or link in three dimensional space is expanded into a sum of $q-$deformed spin networks, and this sum contains topological 
information about the knot and about three-dimensional manifolds obtained by surgery on the knot.  In this way, relatively small
spin networks contain information about the topology of three dimensional spaces. And in this sense one can think of an embedded network
with its woven topology as encoding the ``genetics" of a three-manifold. With this idea in mind, view Figure 13.
\bigbreak

In Figure 13 we illustrate the Borrommean rings, but think of the rings as a network. Then a preparation and result of measurement is illustrated
by cutting one of the components in the rings and inserting a bra and a ket. The underlying network corresponding to the Borrommean rings
can be a spin network expansion as described above, or it can be the result of associating to each of the crossings in the link a
unitary solution to the Yang-Baxter equation \cite{TEQE,Spie,Dye}. If we choose the latter interpretation, then, with an appropriate choice of that 
solution to the Yang-Baxter equation, the trace evaluation of the network can be a topological invariant of the rings. In the case of the
new trace evaluation after preparation and measurement, it will be an invariant of topological movements of the rings that do not
carry strands across the inserted ket or bra. One does not get the luxury of simply removing the segment that is cut. 
\bigbreak

{\it In this way, we see that there is a way to associate the act of cutting a component of a link with an act of quantum measurement.}
The resulting amplitude is {\it not} just the result of removing that component as in the Aravind Hypothesis. From this point of
view the Aravind Hypothesis appears as a radical approximation to the operation of a network model for quantum events. 
\bigbreak

\begin{center}
{\tt    \setlength{\unitlength}{0.92pt}
\begin{picture}(322,207)
\thicklines   \put(151,189){\line(0,-1){114}}
              \put(70,190){\line(1,0){81}}
              \put(131,74){\line(1,0){19}}
              \put(70,74){\line(1,0){50}}
              \put(70,127){\line(0,-1){53}}
              \put(70,190){\line(0,-1){51}}
              \put(125,135){\line(-1,0){11}}
              \put(125,55){\line(0,1){80}}
              \put(12,54){\line(1,0){113}}
              \put(12,134){\line(0,-1){80}}
              \put(22,134){\line(-1,0){10}}
              \put(37,134){\line(1,0){62}}
              \put(30,94){\line(0,1){79}}
              \put(62,94){\line(-1,0){32}}
              \put(108,94){\line(-1,0){34}}
              \put(110,174){\line(0,-1){80}}
              \put(78,174){\line(1,0){32}}
              \put(30,174){\line(1,0){32}}
\thinlines    \put(259,26){$b$}
              \put(201,26){$a$}
\thicklines   \put(189,179){\line(1,0){32}}
              \put(237,179){\line(1,0){32}}
              \put(269,179){\line(0,-1){80}}
              \put(267,99){\line(-1,0){34}}
              \put(221,99){\line(-1,0){32}}
              \put(189,99){\line(0,1){79}}
              \put(196,139){\line(1,0){62}}
              \put(181,139){\line(-1,0){10}}
              \put(284,140){\line(-1,0){11}}
              \put(229,195){\line(0,-1){51}}
              \put(229,132){\line(0,-1){53}}
              \put(229,79){\line(1,0){50}}
              \put(290,79){\line(1,0){19}}
              \put(229,195){\line(1,0){81}}
              \put(310,194){\line(0,-1){114}}
              \put(172,139){\line(0,-1){107}}
              \put(285,140){\line(0,-1){109}}
              \put(173,31){\line(1,0){17}}
              \put(268,29){\line(1,0){17}}
              \put(192,47){\line(0,-1){33}}
              \put(267,46){\line(0,-1){33}}
              \put(193,48){\line(5,-4){23}}
              \put(244,30){\line(5,-4){23}}
              \put(217,30){\line(-3,-2){24}}
              \put(244,31){\line(3,2){23}}
\end{picture}}

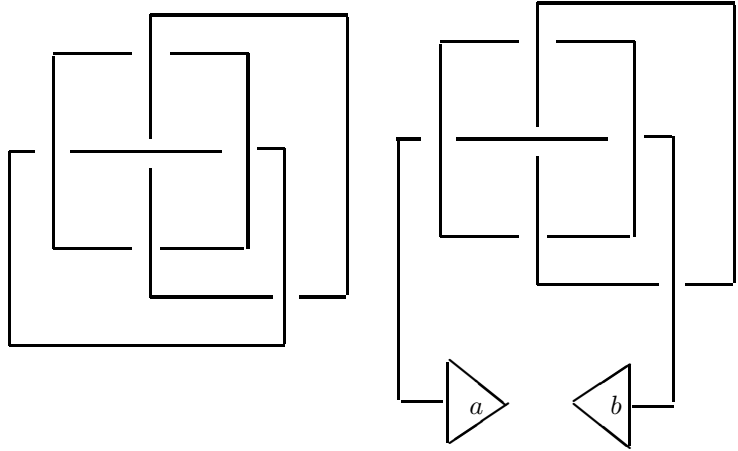
{\bf Figure 13 -  Knotwork - Link Diagram as Quantum Network}
\end{center}

\section{\bf Discussion}
We state some general properties of this quest for relationship between
topology and quantum mechanics: It is normally assumed that one is given the background space over which quantum mechanics appears.
In fact, it is the already given nature of this space that can make non-locality appear mysterious. In writing $|\phi\rangle =
(|01\rangle + |10\rangle)/\sqrt{2},$ we indicate the  entangled nature of this quantum state without giving any hint about the spatial
separation of the qubits that generate the first and second factors of the tensor product for the state. This split between the
properties of the background space and properties of the quantum states is an artifact of the rarefied form given to the algebraic
description of states, but it also indicates that it is the separation properties of the topology on the background
space that are implicated in a discussion of non-locality.
\bigbreak

Einstein, Podolsky and Rosen might have argued that if two points in
space are separated by disjoint open sets containing them, then they should behave as though physically independent. Such a postulate
of locality is really a postulate about the relationship of quantum mechanics to the topology of the background space.  
\bigbreak
 
Approaches such as Roger Penrose's spin networks and the more recent work of John Baez, John Barrett, Louis Crane,
Lee Smolin, Fotini Markoupoulou and others
suggest that spacetime structure should emerge from networks of quantum interactions occurring in a pregeometric, or process phase of 
physicality. In such a spin network model, there would be no separation between topological properties and quantum properties.
\bigbreak

The spin network level is already active in topological models
such as the Jones polynomial, the so-called quantum invariants of knots, links and three-manifolds, topological quantum 
field theories \cite{AT,WIT}, and related anyonic models for quantum computing \cite{F,FR98,FLZ,Freedman5,Freedman6}. For example, the bracket model
\cite{KA87,KA89,KP,QCJP}  for the Jones polynomial can be realized by a generalization of the Penrose $SU(2)$ spin nets to the
quantum group $SU(2)_q.$ 
\bigbreak

In this paper we have placed the Aravind Hypothesis in the larger context of network models for quantum processes. In that context we
can begin  to see why there is something compelling about the hypothesis, even though it is flawed in a multiplicity of ways. In the network model a preparation
and result of measurement is modeled by cutting a graphical edge of the network and inserting a density matrix (a bra and a ket). This operation is remarkably
close to  the Aravind move of cutting a component of a link. The comparison needs further study.
\bigbreak

This paper began with the general notion of a quantum knot as a superposition of classical knots. In this framework the measurment of a quantum knot produces 
a classical knot. We have seen that other notions of measurement lead to tantalizing and sometimes contradictory possibilities. It has been our intent to provide
a wider context for this discussion. At base, the discussion is fundamental. If the world is quantum, then there must be a dialogue that interweaves every
aspect of the apparently classical with the forms of superposition and quantum measurement. We would like to replace space(time) by a weave ${\cal W}$, replace the
weave by  a superposition of weaves ${\cal Q(W)}$ and hence take the world itself as a quantum knot ${\cal Q(W)}$. 

\section*{ACKNOWLEDGMENTS}  
Most of this effort was sponsored by the Defense
Advanced Research Projects Agency (DARPA) and Air Force Research Laboratory, Air
Force Materiel Command, USAF, under agreement F30602-01-2-05022. Some of this
effort was also sponsored by the National Institute for Standards and Technology
(NIST). The U.S. Government is authorized to reproduce and distribute reprints
for Government purposes notwithstanding any copyright annotations thereon. The
views and conclusions contained herein are those of the authors and should not be
interpreted as necessarily representing the official policies or endorsements,
either expressed or implied, of the Defense Advanced Research Projects Agency,
the Air Force Research Laboratory, or the U.S. Government. (Copyright 2004.) It
gives the first author great pleasure to thank Tom Etter, Pierre Noyes, Fernando Souza and Heather Dye for many 
conversations, and the University of Waterloo and the Perimeter Institute for hospitality 
in the course of preparing this paper. \bigbreak

 \end{document}